\documentclass[12pt,preprint]{aastex}

\newcommand{\be} {\begin{equation}}

\newcommand{\ee} {\end{equation}}

\def\src  {RX\,J0806.3+1527}

\newcommand{\VLT}{{\VLT} }

\newcommand{\bc}{\begin{center}}
\newcommand{\ec}{\end{center}}
\def\ltsima{$\; \buildrel < \over \sim \;$}
\def\lsim{\lower.5ex\hbox{\ltsima}}
\def\loe{\lower.5ex\hbox{\ltsima}}
\def\gtsima{$\; \buildrel > \over \sim \;$}
\def\gsim{\lower.5ex\hbox{\gtsima}}
\def\goe{\lower.5ex\hbox{\gtsima}}
\newcommand {\rc}{\rm}

\shortauthors{ISRAEL ET AL.}
\shorttitle{Unveiling the Nature of \src}

\begin{document}
\title{The glitches of the Anomalous X--ray Pulsar 
\newline 
1RXS J170849.0--400910}

\author{
S.~Dall'Osso\altaffilmark{1,2}
G.~L.~Israel\altaffilmark{2,3}, 
L.~Stella\altaffilmark{2,3},
A.~Possenti\altaffilmark{4,5},
E.~Perozzi\altaffilmark{6}
}

\altaffiltext{1}{Universit\`a degli Studi di Roma ``La Sapienza'', P.zzle Aldo 
Moro 5, I-00195 Roma, Italy}

\altaffiltext{2}{INAF - Osservatorio Astronomico di Roma, V.~Frascati 33, 
       I-00040 Monteporzio Catone (Roma), 
       Italy; dallosso, gianluca and stella@mporzio.astro.it}

\altaffiltext{3}{Affiliated to the International Center for Relativistic 
Astrophysics}

\altaffiltext{4} {INAF - Osservatorio Astronomico di Bologna, via Ranzani 1, 
40127 Bologna, Italy}

\altaffiltext{5}{INAF - Osservatorio Astronomico di Cagliari, Loc. Poggio dei 
Pini, Strada 54, 09012 Capoterra (CA), Italy; possenti@ca.astro.it}

\altaffiltext{6}{Telespazio, Via Tiburtina 965, 00156 Roma, Italy; 
Ettore\_Perozzi@telespazio.it}

\thispagestyle{empty}

\begin{abstract} 
We report on a timing analysis of archival observations of the 
Anomalous X--ray Pulsar 1RXS J170849.0--400910 made with the RXTE Proportional 
Counter Array. We detect a new large glitch 
($\Delta \nu / \nu \simeq 3\times 10^{-6}$) which occurred between 2001 March 
27 and 2001 May 6, with an associated large increase in the spin--down rate 
($\Delta \dot{\nu}/\dot{\nu}\simeq$ 0.3). 
The short time (1.5 yrs) elapsed from the previously detected glitch 
and the large amplitude of the new spin--up place this source among the most 
frequent glitchers, with large average glitch amplitudes, similar to those of 
the Vela pulsar. 
The source shows different recoveries after the glitches: in the first one it
is well described by a long term linear trend similar {\rc to} those seen in  
Vela--like glitches; in the second case the recovery is considerably faster
and is better described by an exponential plus a fractional change in the 
long--term spin--down rate of the order of 1\%. No recovery of the latter 
is detected but additional observations are necessary to confirm this result. 
We find minor but significant changes in the average pulse profile after both 
glitches. No bursts were detected in any lightcurve, but our search was 
limited in sensitivity with respect to short (t$<$ 60 ms) bursts. 
Observed glitch properties are compared to those of radio pulsar glitches;
current models are discussed in light of our results. It appears that 
glitches may represent yet another peculiarity of AXPs. Starquake--based 
models appear to be prefered on qualitative grounds. Alternative models can be 
applied to individual glitches but fail in explaining both. Thus the two 
events may as well arise from two different mechanisms.
\end{abstract}


\section{Intoduction}

AXPs are a small class of X--ray pulsars whose luminosity, though somewhat
lower ($10^{34}$--$10^{36} \mbox{erg s}^{-1}$) and much softer than that
of standard X--ray pulsars, largely exceeds their rotational energy losses
($\sim 10^{32}$-- $10^{33}\mbox{erg s}^{-1}$). The association with systems 
accreting matter at low level (Mereghetti \& Stella 1995, Van Paradijs et 
al. 1995, Chatterjee et al. 2000, Alpar 2001)
is made problematic by some of the observational properties that ultimately 
characterize AXPs as a distinct class.
These are their fairly steady spin--down, clustering in a narrow range of spin
periods (6--12 sec) and absence of conspicuous counterparts
at other wavelengths (Mereghetti \& Stella 1995, Mereghetti et al. 2002).
For three of these objects very weak infrared counterparts have so far been 
identified (Hulleman et al. 2001, Wang \& Chakrabarty 2002, 
Israel et al. 2002, Israel et al. 2003). In one case an optical counterpart, 
with pulsations at the X--ray period, was found (Kern \& Martin 2002).
Though faint, these optical/IR counterparts are far too bright with respect to
the extrapolation of a blackbody--like spectrum matching the X--ray emission. 
The origin of the optical/IR emission of AXPs remains to be understood 
(Israel et al. 2003).

Furthermore, both enhancements of the noise level and short--duration bursts 
in the X--ray emission of some sources have been found after years 
of monitoring (Kaspi et al. 2001, Gavriil and Kaspi 2002a) and
evidence for a spin--up event consistent with a radio pulsar--like
 glitch has been found in RXS J170849.0--400910 (Kaspi et al.2000, Gavriil and 
Kaspi 2002b). This lent 
additional support to the idea that AXPs may be magnetars, isolated neutron 
stars endowed with a huge magnetic field ($B > 3 \times 10^{14}$ G) whose
energy provides the reservoir to power the X-ray luminosity.
This would link them to SGRs, a small class of bursting X--ray sources sharing
 with AXPs a comparable quiescent X--ray luminosity  as well as
similar periods and spin--down rates. For SGRs, several independent arguments 
hint to the presence of extremely high magnetic fields 
(Thompson \& Duncan 1995).
Very recently this picture has received compelling evidence by the detection 
of an SGR--like X--ray outburst in 1E 2259+586 (Kaspi et al. 2003). 
Associated with the outburst a glitch was also detected, with a large 
amplitude in the rotation rate ($\Delta \nu / \nu \sim 4\times 10^{-6}$) and 
extreme amplitude in the spin--down rate ($\Delta \dot{\nu} / \dot{\nu} 
\sim 1$). Remarkably, no such direct connection has ever been observed, even 
in the case of SGR bursts (Woods et al. 2003, 2002), apart indirect 
evidence in one case (Woods et al.1999). 
   
Given the previous detection of a glitch in RXS J170849.0--400910, we decided 
to analyse timing data following the glitch since the study of the long--term 
recovery of the source, including the possible occurrence of new glitches 
within a couple of years, would allow a closer comparison with the 
properties of known glitching radio pulsars. Indeed, glitches in 'adolescent' 
radio pulsars ($t_s\sim 10^3\div10^5$ yrs) repeat at intervals of a few years 
or less, while many of the youngest and oldest radio pulsars have much sparser
glitches, if at all. In the magnetar model the influence of the magnetic field
on the structure and dynamical evolution of a neutron star is expected to 
produce features that may be peculiar to AXPs and can be identified by 
precision timing.

\section{Data selection}

We analyzed data from archival observations of RXS J170849.0--400910 carried 
out with the RXTE Proportional Counter Array (PCA). The total data 
set spans more than 4 years, between 1998 January 13 and 2002 May 29.

The PCA is made up of five PCUs each split in two, an upper propane 
layer and the main Xenon volume through which run 3 different Xenon layers.
We analyzed data in the GoodXenonwithPropane configuration, in which
single events, each coming from one of the 3 Xenon layers of one of the PCUs, 
are recorded in sequence at 1$\mu$s time resolution. 
Events are analyzed and recorded by two independent Event Analyzers (EAs) at 
the same time.  
 
Following Kaspi et al. (2000), in order to maximize the sensitivity, the 
analysis was restricted to unrejected events in the top Xenon layer of each 
PCU and only soft photons were selected for extracting lightcurves, namely 
those collected in channels 6--14 (corresponding to the range $\sim 
2.5-6$keV). 

Data from the two EAs were merged into a single lightcurve using the standard 
RXTE software and then rebinned at $62.5$ ms resolution. Standard reduction to
barycentric dynamical time (TDB) was then applied, using 
the (10$''$ accurate) source position given by Israel et al. (1999).

\section{Data Analysis}

A phase--coherent timing analysis was carried out using a phase--fitting 
tecnique, an iterative procedure allowing the determination of the
 frequency of a coherent signal \textit{at a given epoch}, with a precision 
that increases as more observations are included in the analysis.

The signal from the source is assumed to be periodic 

\begin{equation}
\label{seriediphi}
\phi(t)=\phi_{0}+\int_{t_0}^{t} \omega(t')dt'
\end{equation}

The steady variation of the period is parametrized by a series of time 
derivatives, truncated at the highest term which appears statistically 
significant. 

After integration equation (\ref{seriediphi}) reads:

\begin{equation}
\label{phasefit}
\phi(t)= \phi_{0}+\omega_{0}(t-t_{0})+ \frac {1}{2} \dot{\omega}_{0}(t-t_{0})^{2}+ \frac{1}{6} \ddot{\omega}_0(t-t_{0})^3+...
\end{equation}
where the $\cdots$ include all the terms in the series resulting from 
derivatives of $\omega$ of order higher than the second.

A first guess period (P$_{0}$) was determined through
an epoch folding search of the lightcurve of an observation chosen as
the starting point.
Its epoch was adopted as the reference with respect to which all the following
phases were calculated.

A template pulse--profile was then determined by folding this 
lightcurve at the period P$_{0}$ and fitting the resulting binned 
pulse--profile with a template function of the form:

\begin{equation}
\label{template}
\psi(t)=\sum_{i=1}^{n} ~A_{i} sin[i \omega_0(t-\psi_{i})]
\end{equation}
truncating the series at the highest armonic that {\rc was}
 statistically significant after performing an F--test at the 99 \%
($ 2.7 \sigma$) level.

The lightcurve was then divided in 4 intervals and each of them folded at the 
same P$_{0}$. The 4 resulting binned pulse--profiles were fit by the 
template function, with A$_1$ and $\psi_1$ as free parameters 
while keeping frozen the harmonics amplitude ratios (A$_i$/A$_1$) and phase
differences $\psi_i- \psi_1$. 
This is equivalent to assuming that the shape of the average pulse profile does
not change. We checked this by verifying that the $\chi^2$s were 
distributed around the expected value $<\chi^2>=\nu$ with a dispersion 
comparable to $\sigma_{\chi^2}=\sqrt{2\nu}$ with $\nu$ the degrees of 
freedom of the model.

The values of $\psi_{1}$ for the 4 intervals were then plotted versus the 
epochs and fitted by a function of the form (\ref{phasefit}), truncating that 
series at the last coefficient which was found to be statistically significant
through an F--test at the 99\% level. A 99\% confidence 
interval for the value of the first parameter consistent with zero was 
also determined.

Phase--fitting applied to the four phases provides a correction to the guess 
period $P_0$, resulting in a better estimate of the period -- call it P$'$ -- 
with a smaller uncertainty. The lightcurve 
was folded again using P$'$, in order to have a more accurate template 
function.

Then we evaluated the maximum time interval $\Delta t$ over which extrapolated
phases remained coherent without any cycle ambiguity. 
Typically $\Delta t \sim $ {\rc 8} hrs when P is determined with a single 
observation {\rc  whose duration is $\sim$ 3 ks}. The observation selected as 
the starting point of our phase coherent analysis had a separation from the 
closest observation smaller than $\Delta t$.

Phase--coherence could then be maintained over the $\sim$ 4 days separating the
first three observations. The addition of the third observation yielded as 
small an uncertainty on P ($\sim$ some $\mu$s) that $\Delta t \sim 40\div 50$ 
days, enough to connect all the following  observations, typically spaced by 
$\sim 30$ days. Of course, the addition of more observations further decreases 
the uncertainties on P.

The fit to the fourth observation, more than 30 days apart from the 
first one, required the inclusion of a significant quadratic term ($\dot{P}$)
in Eq. (\ref{phasefit}).

From this point onwards the phase--coherent analysis was further extended
by using the $\dot{P}$ derived from the fits together with $P$, 
readjusting the values of both parameters at each step, checking for 
consistency with previous error intervals and readjusting the template
function, that was always obtained after folding at the best $P$ and $\dot{P}$
all the lightcurves that had been connected with a single solution of equation
(\ref{phasefit}).

By this procedure we analyzed all observations since January 13 1998 
(MJD 50825). We detected the glitch already reported by Kaspi et al. (2000) 
between 1999 September 25 and 1999 October 21, as our timing solution failed 
to match the observed phases later than September 25 even with the addition 
of higher orders in the polynomial expansion of Eq. \ref{phasefit}. The spin 
parameters previous and after the glitch and the inferred glitch parameters 
are reported in Tab. \ref{glitch1}. We note that our results are in agreement 
with those of Kaspi et al. (2000) and Gavriil \& Kaspi (2002b).

\begin{figure*}[p] 
\plotone{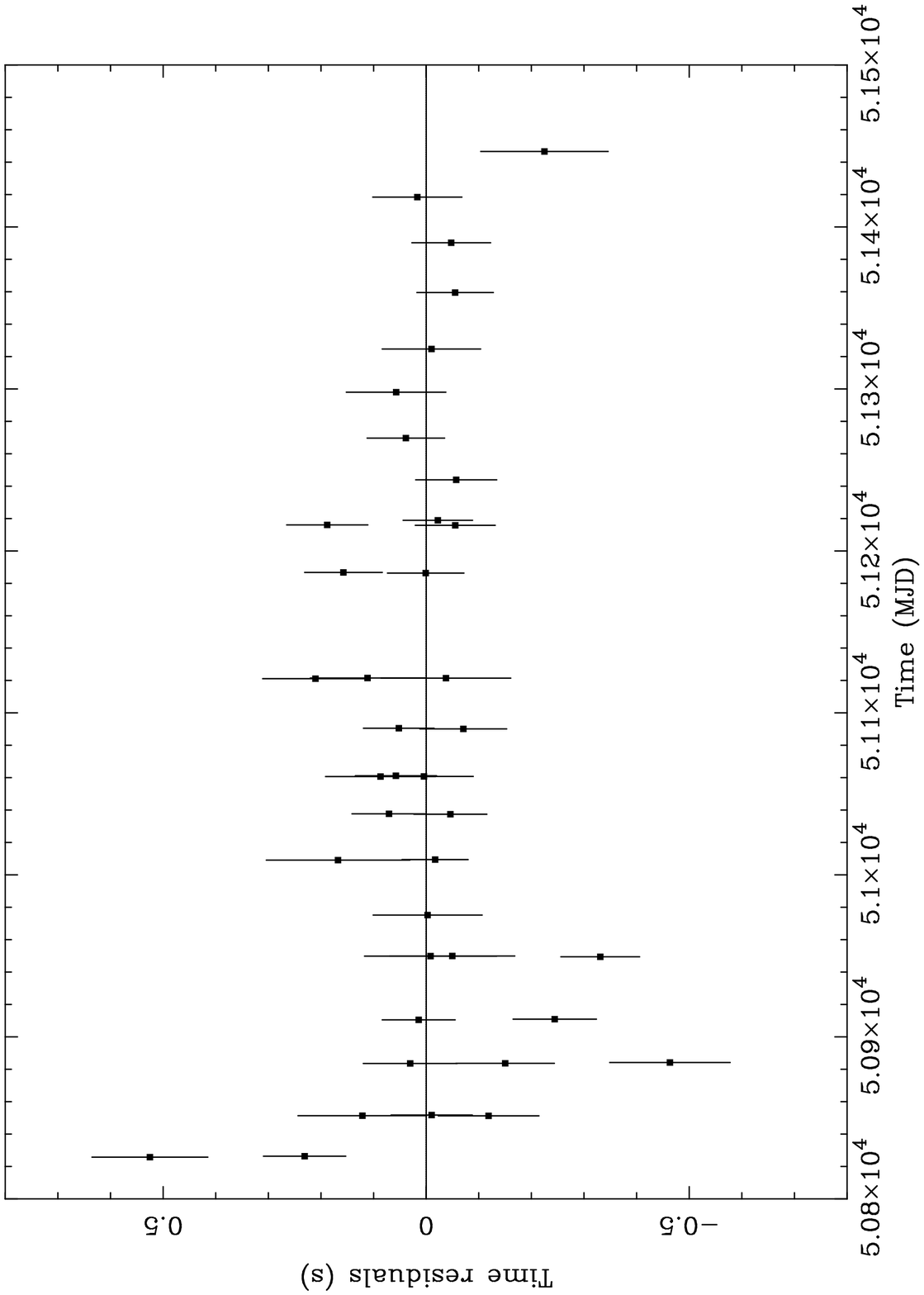}
\caption{}
\label{preglitch1}
\end{figure*}

Time residuals after subtraction of our pre--glitch model are shown in
Fig.\ref{preglitch1}.

\section{New Data and Results}
\label{nuovidati}

The set of new data started with {\rc 4} closely spaced observations {\rc , 
one} made on 2001 May 6, {\rc two on} May 11 and {\rc one on} May 12 
(MJD 52035, 52040 and 52041, respectively). 

\begin{figure*}[p] 
\plotone{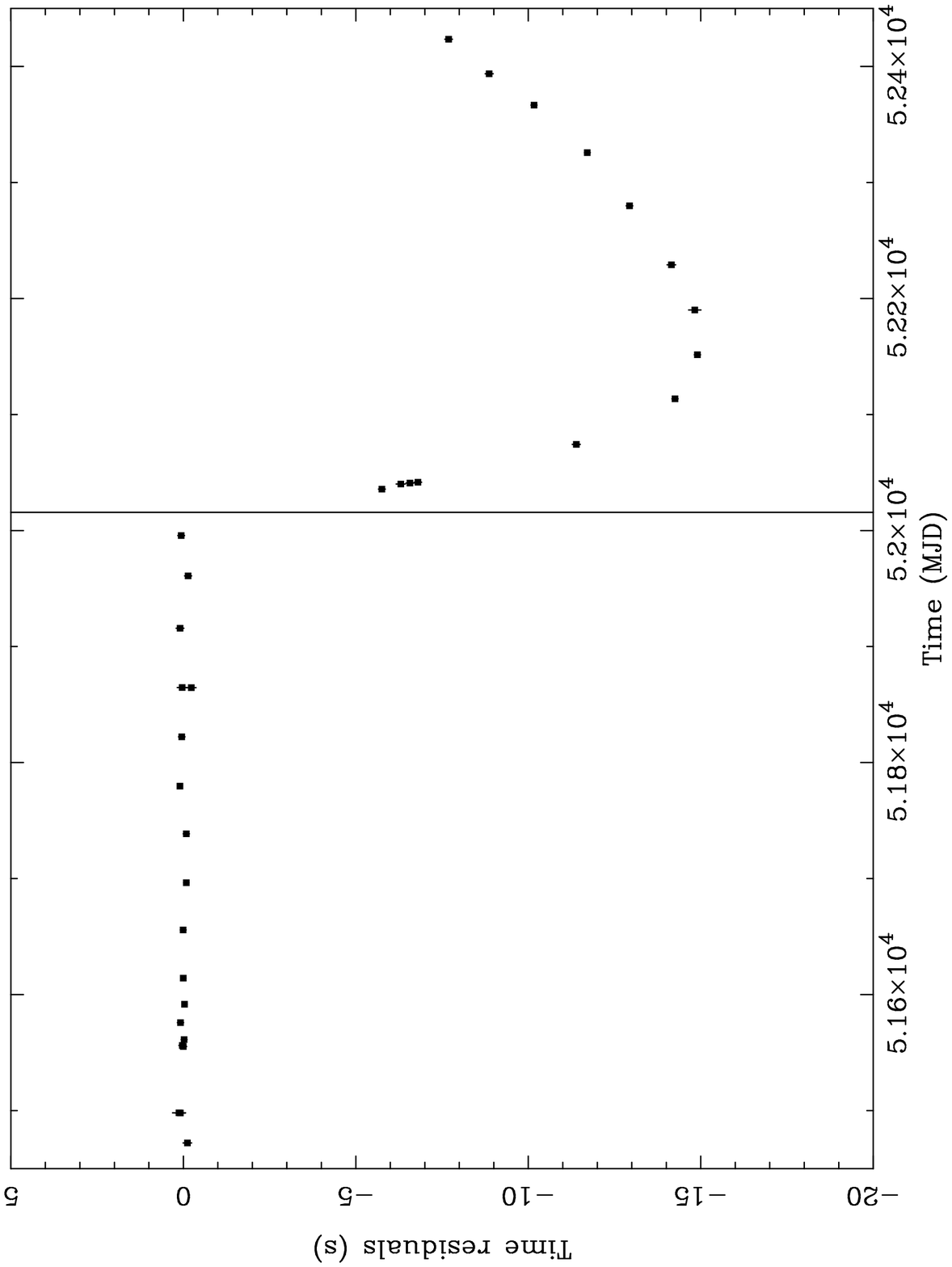}
\caption{}
\label{aprilglitch}
\end{figure*}

Fitting these observations with post--glitch N.1 model proved impossible: this 
is apparent in Fig.\ref{aprilglitch}, where the timing residuals after 
subtraction of the model for recovery from glitch N.1 are shown and the sudden 
change in the star spin is clearly seen.

This feature is qualitatively similar to the 1999 glitch 
(see Kaspi et al. 2000), suggesting that another glitch may have happened 
between March 27 and May 6, 2001.

We then repeated the phase--fitting on the series of data from 2001 May 6 to
2002 May 29, taking as the reference epoch the middle point between {\rc the 
second} May 11 and {\rc the} May 12 observations.
The {\rc latter} had a longer duration ({\rc $\sim$ 2.5 ks}) so we chose it as 
the starting point of a new phase coherent solution, in order to get as small 
an uncertainty on P$_0$ as possible.

The separation between {\rc these} two observations was too long 
($\sim$ {\rc 18} hrs) for coherence to be maintained and there was an 
ambiguity of 1 cycle in the phases of the second observation due to a $\sim$ 
3 ms uncertainty on P$_{0}$ determined from the first one only. 
A coherent solution connecting the {\rc four} observations 
between May 6 and May 12 could be found only assuming that a whole cycle 
had been missed from the first to the second one using that value of P$_{0}$. 
{\rc With} this assumption the fit to the {\rc 4} observations was obtained 
with no significant period derivative. 

The new P {\rc was} $\sim 20 \mu$s shorter than expected from extrapolation
of the pre--glitch model and, when a {\rc fifth} observation was added, a 
$\dot{P}$ term was highly significant with a value greater than that 
measured before May 6. This is consistent with the positive $\Delta \dot{P}$ 
expected from a glitch.

\begin{figure*}[p] 
\plotone{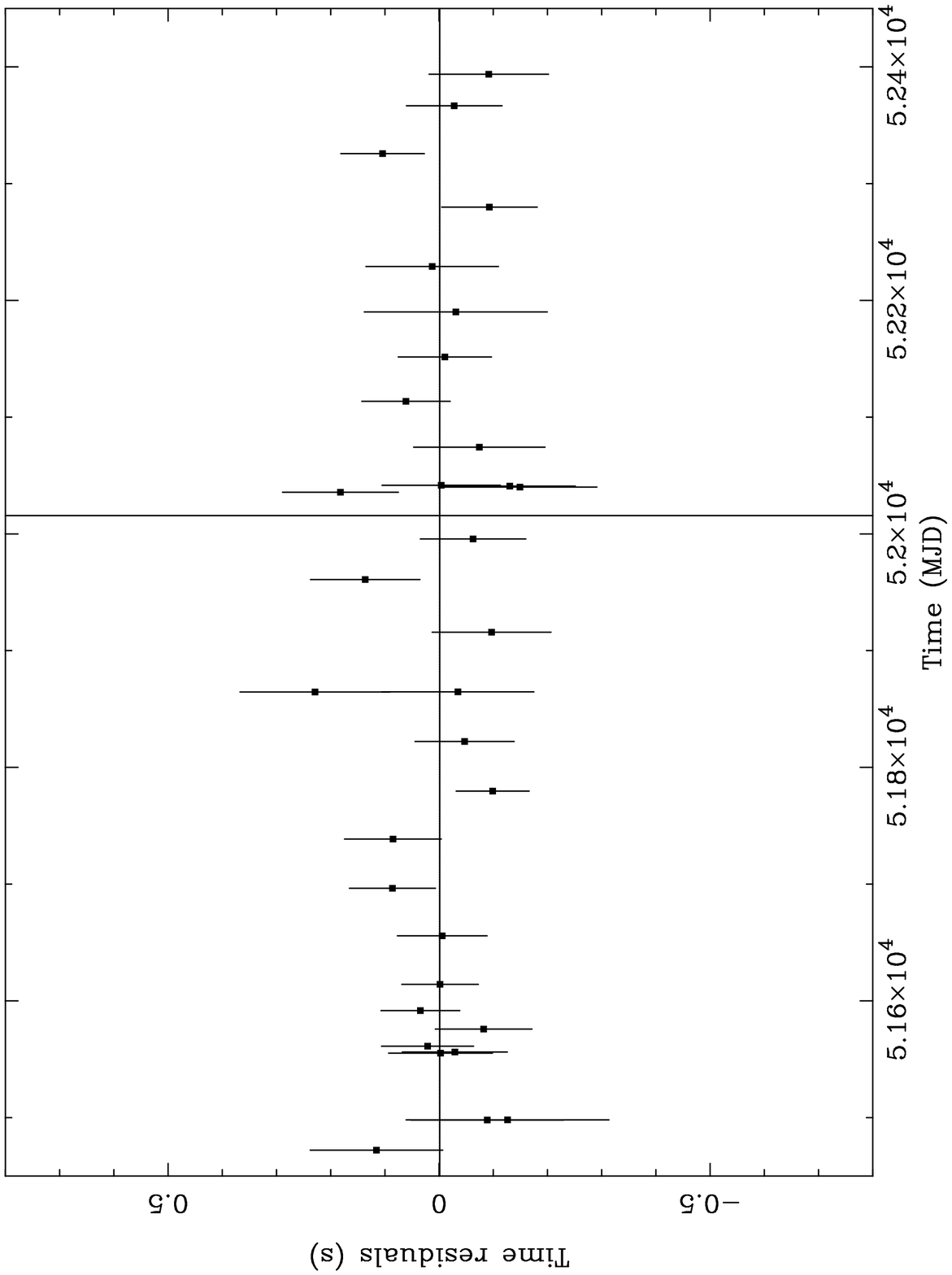}
\caption{}
\label{residuiprepost}
\end{figure*}

The parameters of the best--fit polynomial solution, up to the fourth frequency
derivative, connecting all observations from 2001 May 6 to 2002 May 29 
(MJD 52423) are reported in Tab.\ref{glitch2} and residuals with respect to it 
are shown at the right of the vertical line in Fig.\ref{residuiprepost}.

We note, however, that the period evolution can be interpreted in a different 
way.
Post--glitch recoveries of radio pulsars are usually described in 
terms of two different components with a long--term, approximately linear 
recovery of $\dot{\nu}$ superposed to an initial, short--term exponential
that, in some cases, has been resolved in the sum of a few exponentials 
with different timescales. The recovery fraction is usually defined as
Q=$\delta \nu_{exp}/ (\delta \nu_{exp}+\Delta \nu_l)$ (Eq.\ref{esp} below).

Thus, given the relatively fast recovery after glitch N.2 and the higher order
polynomial coefficients required, we also tried fitting the phases following
this glitch with the sum of an exponential and a linear trend. 
The same procedure was applied for consistency to the recovery from glitch 
N.1 as well. 
To this aim, we folded the lightcurves with the pre--glitch values of $\nu$ 
and $\dot{\nu}$, obtaining a plot of the phase residuals with respect to the 
extrapolation of the pre--glitch solution ($\nu_0$ below). The fit was then
made through the formula:

\begin{equation}
\label{esp}
\nu(t)- \nu_0(t)= \delta \nu_{exp} \mbox{~exp} \left(-\frac{t-t_g}{\tau}\right) + \Delta \nu_l + \Delta \dot{\nu}_l(t-t_g) + (1/2) \Delta \ddot{\nu}_l 
(t-t_g)^2
\end{equation}

with $\phi(t)=\phi(t_g) + 2\pi\int_{t_g}^{t} [\nu(t')-\nu_0(t')] dt'$

The epoch of the glitch ($t_g$) was fixed at the midpoint between the last 
pre--glitch and the first post--glitch observation.  

The parameters of the best--fit exponential solution to glitch N.2 are also 
reported in Tab.\ref{glitch2}; these include a significant change in 
$\dot{\nu}$ with respect to the pre--glitch solution while both $\Delta \nu_l$
and $\Delta \ddot{\nu}_l$ are fixed to zero. This because no significant 
improvement of the fit is obtained allowing also for changes of either 
parameters. However we report in Tab. \ref{glitch2} a 99.9\% confidence 
interval for a possible $\Delta \nu_l$. 

In glitch N.1 a good fit is obtained with either a step in $\nu$ or in 
$\dot{\nu}$, while allowing for both does not prove statistically significant 
at the 99\% confidence level. 

Based on the $\chi^2$ (or the r.m.s.) of the fits, the exponential recovery 
solutions do not provide a preferable description of the observations. 
However, at glitch N.2 the exponential model fits the evolution of the observed
phases with a smaller set of parameters with respect to the polynomial 
expansion.

{\rc We note that very recently Kaspi \& Gavriil (2003) have carried out a
similar timing analysis of this source detecting glitch N.2 as well. Their 
results concerning this glitch are somewhat different from ours, since they 
assumed a step $\Delta \nu_l \sim 1.3\times 10^{-8} \mbox{s}^{-1}$ superposed 
to the exponential component and no significant change in $\dot{\nu}$. As they 
suggest, this apparent discrepancy is due to the non--inclusion of the 
$\ddot{\nu}$--term in their fit during the interglitch time, since this may be 
contaminated by timing noise. However the post--glitch N.1 
$\Delta \dot{\nu}/\dot{\nu}$ they find is wholly unrecovered and amounts 
almost exactly to the unrecovered $\Delta \dot{\nu}/\dot{\nu}$ 
($\sim \mbox{10}^{-2}$) we found after glitch N.2. On the other hand, 
$\dot{\nu}$ in our model for glitch N.1 is completely recovered and returned 
to its pre--glitch value right before the second glitch. Thus, the apparently 
different solutions do indeed agree in that the long--term  $\dot{\nu}$ after 
the second glitch is greater by $\sim$ 1$\%$ than its value previous to the 
first glitch (see Fig. 1, bottom panel of Kaspi \& Gavriil 2003).

Concerning the nature of the interglitch $\ddot{\nu}$, we stress that its
inclusion in the post--glitch N.1 model provides a great improvement of 
the fit to the phases. The $\chi^2$ value changes from 72 (with 16 d.o.f.) 
to 11 (with 15 d.o.f.), which corresponds to a significance level greater than 
5$\sigma$. In the solution previous to glitch N.1 the significance of 
$\ddot{\nu}$ was at the 2.6 $\sigma$ level if the first data point, which 
deviated significantly from the average of all other points, was excluded from 
the fit, and at less than 4$\sigma$ in any case (see Kaspi et al. 2000).
From Fig. 2 of Kaspi \& Gavriil (2003) the residuals with respect to their 
fits appear to increase with time, somewhat in connection with the occurrence 
of glitches, while we do not find any hints to such variations.}

\begin{figure*}[p] 
\plotone{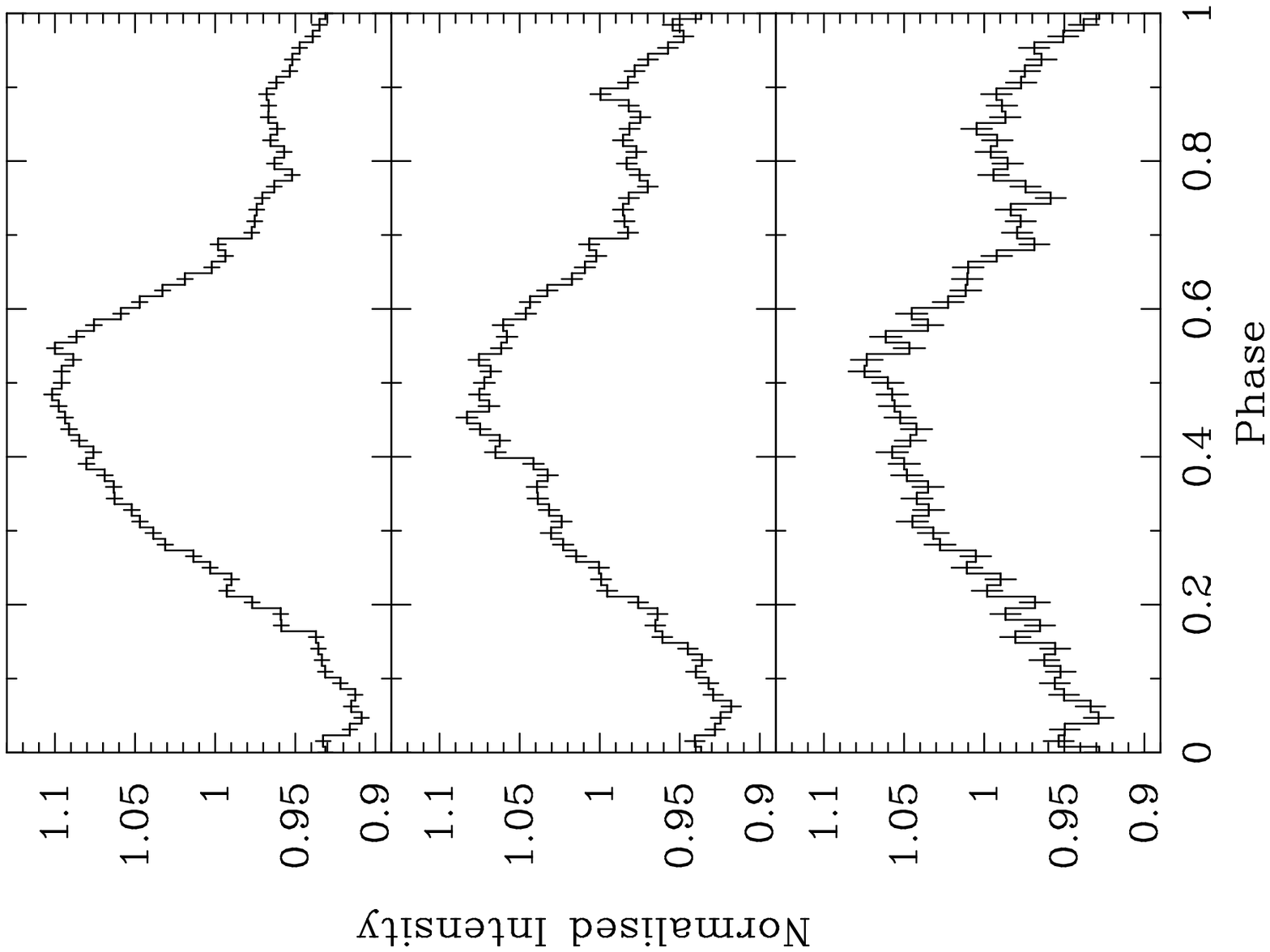}
\caption{}
\label{pulse}
\end{figure*}

After both glitches we found small but statistically significant changes in 
the shape of the average pulse profile. The templates for each of the three 
intervals are shown in Fig. \ref{pulse}. Each of them was obtained 
after folding a reduced set of observations, namely only those in the
corresponding interval that could be phase--connected with a simple 
($\nu, \dot{\nu}$) model. The adopted values of $\nu$ and $\dot{\nu}$ for the 
folding were those derived by the best--fit to this reduced set, instead of 
the global best--fit parameters. Though small, observed changes are 
statistically significant. 
The best--fit parameters for the three average pulse profiles are reported in 
Tab. \ref{pulsemedio}. 
After glitch N.1 the average pulse--profile can still be fit with three 
harmonics, but changes in the amplitude ratios and phase 
differences are 3--5 times greater than reported 1$\sigma$ errors. On the 
other hand, after glitch N.2 a fourth harmonic becomes statistically 
significant ($\chi^2$ = 69.1 with 59 d.o.f. and $\chi^2$ = 49.0 with 57 
d.o.f.) while the amplitude ratios and phase differences of the first three 
components are consistent with those determined previous to the glitch.

Short--duration bursts in the X--ray emission of two AXPs have been recently
detected (Kaspi et al. 2002, Kaspi et al. 2003). Since the recently detected
outburst in 1E2259+586 was also associated to a large glitch, we looked 
for bursts in the same X--ray lightcurves used for the phase--fitting. These 
were binned at 62 ms and only photons in the 2.5--6 keV range were selected, 
so our search was somewhat limited in sensitivity. 
Each lightcurve was searched for fluctuations above a significance 
threshold of 8$\sigma$ with respect to its mean count rate. No events were
found above the threshold, so our data do not show evidence for bursts. Given 
the limitations discussed above the possibility of very short--term 
($t<60$ ms) bursts cannot be ruled out.

\section{A comparison of the two detected glitches}

The most striking feature of the glitching behaviour of this source is the 
different appearence of the two spin--up events, in particular the 
remarkably different changes in $\dot{\nu}$ and in the recovery timescales, 
both unusual findings compared to radio pulsars.

Following glitch N.1 the first derivative of the spin frequency ($\dot{\nu}$) 
went back to its pre--glitch value at a constant rate (constant 
$\ddot{\nu}$), over one year or so, with no significant exponential short--term
 component. 
Both the inferred changes $\Delta \nu/\nu$ and $\Delta\dot{\nu}/\dot{\nu}$ are 
in the range of those observed in Vela--like glitches, where tipically 
$\Delta\nu / \nu \sim 10^{-7}\div 10^{-6}$ and $\Delta\dot{\nu}/\dot{\nu} 
\sim 10^{-3}\div10^{-2}$.

The recovery time $\Delta t$ can be evaluated setting to zero the expression 
of the frequency difference between the post--glitch and the extrapolated 
pre--glitch solution:
\begin{equation}
\label{deltanu}
\Delta \nu(t)=\Delta \nu_{g}+ \Delta \dot{\nu}_g (t-t_g) + \frac{1}{2} \Delta \ddot{\nu}_g(t-t_g)^{2}
\end{equation}
where the subscript $g$ indicates quantities evaluated at the assumed epoch of 
the glitch, $t_g$. It obtains $\Delta t \sim$ 400 d, to be
compared with the $\sim$ 600 d one obtains for the recovery of 
$\Delta\dot{\nu}_g$. This discrepancy is not very significant however, given 
the uncertainty on $\ddot{\nu}$ and on the extrapolated value of $\nu$. 
It can be concluded that $\nu$ and $\dot{\nu}$ have approximately relaxed back 
to their 
pre--glitch values over a timescale $\sim$ 400--600 d. The constant 
second derivative measured after glitch N.1 may then represent a long--term 
linear recovery of $\dot{\nu}$ towards its preglitch value. 
In this case, it is characterized by a timescale comparable -- though somewhat 
shorter -- to typical values ($>$ 2--3 yrs) of radio pulsars with the same 
linear response.

Glitch N.2 is, on the other hand, quite different. The fractional spin--up 
$\Delta \nu / \nu \sim 3 \times 10^{-6}$ is comparable to the giant glitches 
observed in the Vela and other pulsars of similar spin--down age and 
$\Delta \dot{\nu} / \dot{\nu}$ is one of the largest observed ever. 
Comparable changes in the spin--down rate have been detected in a few giant 
glitches of the Vela pulsar (Lyne et al. 2000), but were followed by a fast 
recovery ($\tau \leq$ 1 day, Flanagan 1990). On the contrary, the recovery 
from glitch 2 of RXS J170849.0--400910 is found to last $\sim$ 100--200 days. 
The only other long lasting ($\sim 60$ d) large change in $\dot{\nu}$
is the one recently observed in association with an X--ray outburst in the 
AXP 1E 2259+586 (Kaspi et al. 2003). 

Residual arrival times of Fig. \ref{aprilglitch} show that, after glitch N.2,
the source evolves towards a rotational period longer than the extrapolation 
of the pre--glitch solution. Indeed, had the recovering $\nu(t)$
asymptotically approached the extrapolated pre--glitch solution from above, 
time residuals would have asymptotically settled towards a constant offset
with respect to pre--glitch ones. This is a 
peculiar result, found rarely in other sources. One remarkable case is 
the Crab glitch of 1975 February 4 where, due to a persistent (at least for 3 
yrs) fractional increase in the spin--down rate of the order $\sim 10^{-4}$, 
the spin frequency after the exponential recovery ($\tau\sim$ 50 d) became 
slower than expected from the pre--glitch solution (Link, Epstein and Baym, 
1992). 
This is an interesting constraint on any model, indeed, and will be discussed 
in Sec. \ref{vortexcreep}.

As already discussed in the previous section, the quick recovery from glitch 
N.2 suggests that a different function, such as the exponential decay of the 
kind given by Eq. (\ref{esp}), may be better suited to fit the observed phases.

However, in order to obtain an acceptably good fit for the recovery form 
glitch N.2, one must allow for a steepening of $\dot{\nu}$ 
($\Delta \dot{\nu}_l/\dot{\nu} \simeq 9\times 10^{-3}$, where the 
denominator is the pre--glitch value of $\dot{\nu}$ reported in Tab. 
\ref{glitch2}) superposed to the exponentially recovered spin--up 
($\chi^2$=18.45 with 10 d.o.f.). Large recovery fractions and permanent 
increases in the spin--down rate are typical of the Crab pulsar glitches, but 
these are usually smaller events, with $\Delta \nu/\nu <10^{-7}$ and 
$\Delta \dot{\nu}_l/\dot{\nu} <10^{-3}$. Thus differences and similarities 
with both Vela--like and Crab--like glitches are found in glitch N.2. 

An alternative fit to the phases, with a step in $\nu$ superposed to the 
exponential component and $\Delta \dot{\nu}_l$ of Eq. (\ref{esp}) fixed to 
zero gave a worse fit ($\chi^2$=30.9 with 10 d.o.f.). Finally, when {\rc both 
$\Delta \nu_l$ and $\Delta \dot{\nu}_l$ were left as free parameters, no 
significant improvement of the fit was obtained ($\chi^2=$17.71 with 9 d.o.f.) 
with respect to the model above with $\Delta \dot{\nu}$ only}.

\begin{figure*}[p] 
\plotone{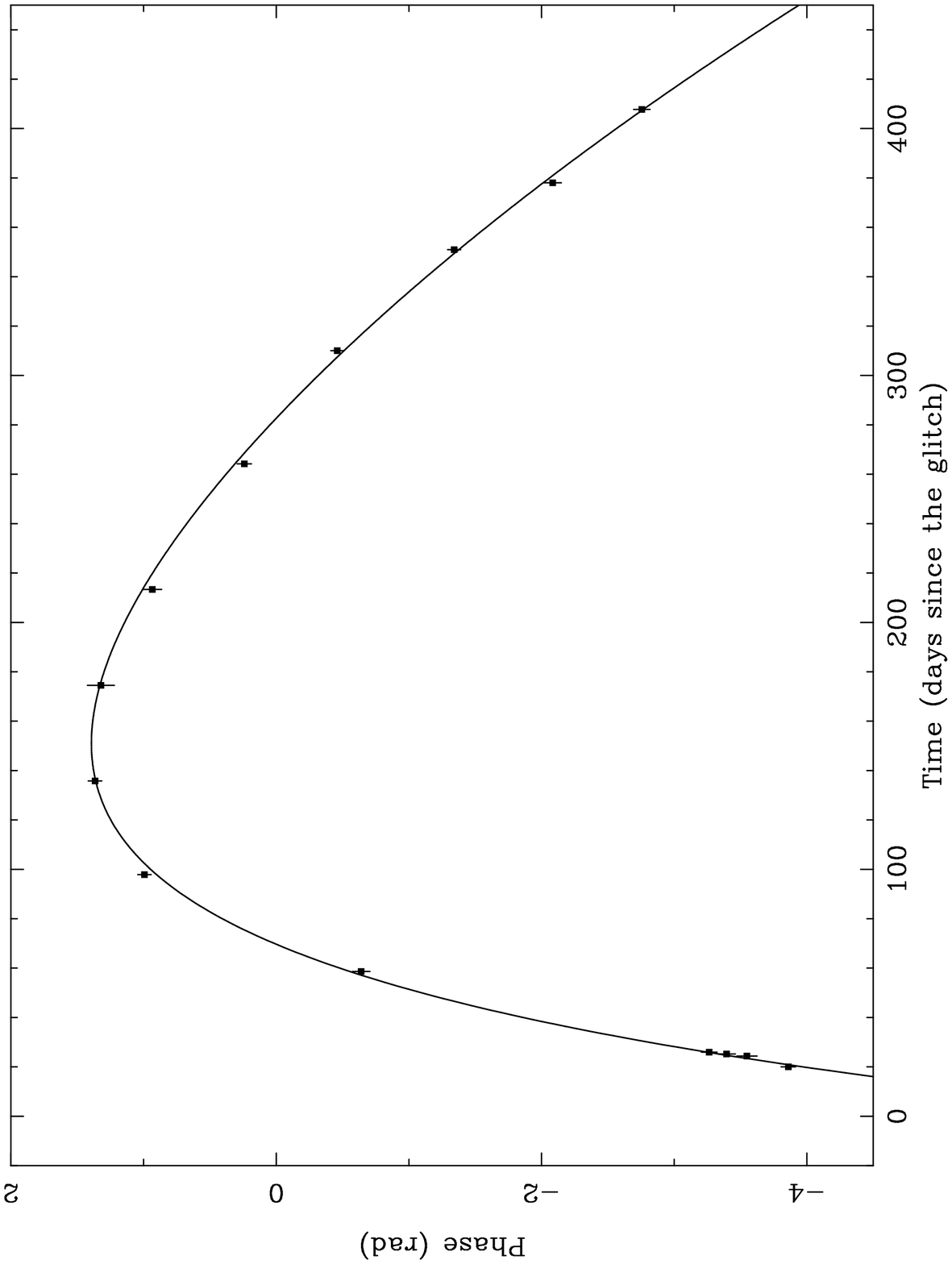}
\caption{}
\label{exp}
\end{figure*}

\begin{figure*}[tbh] 
\plotone{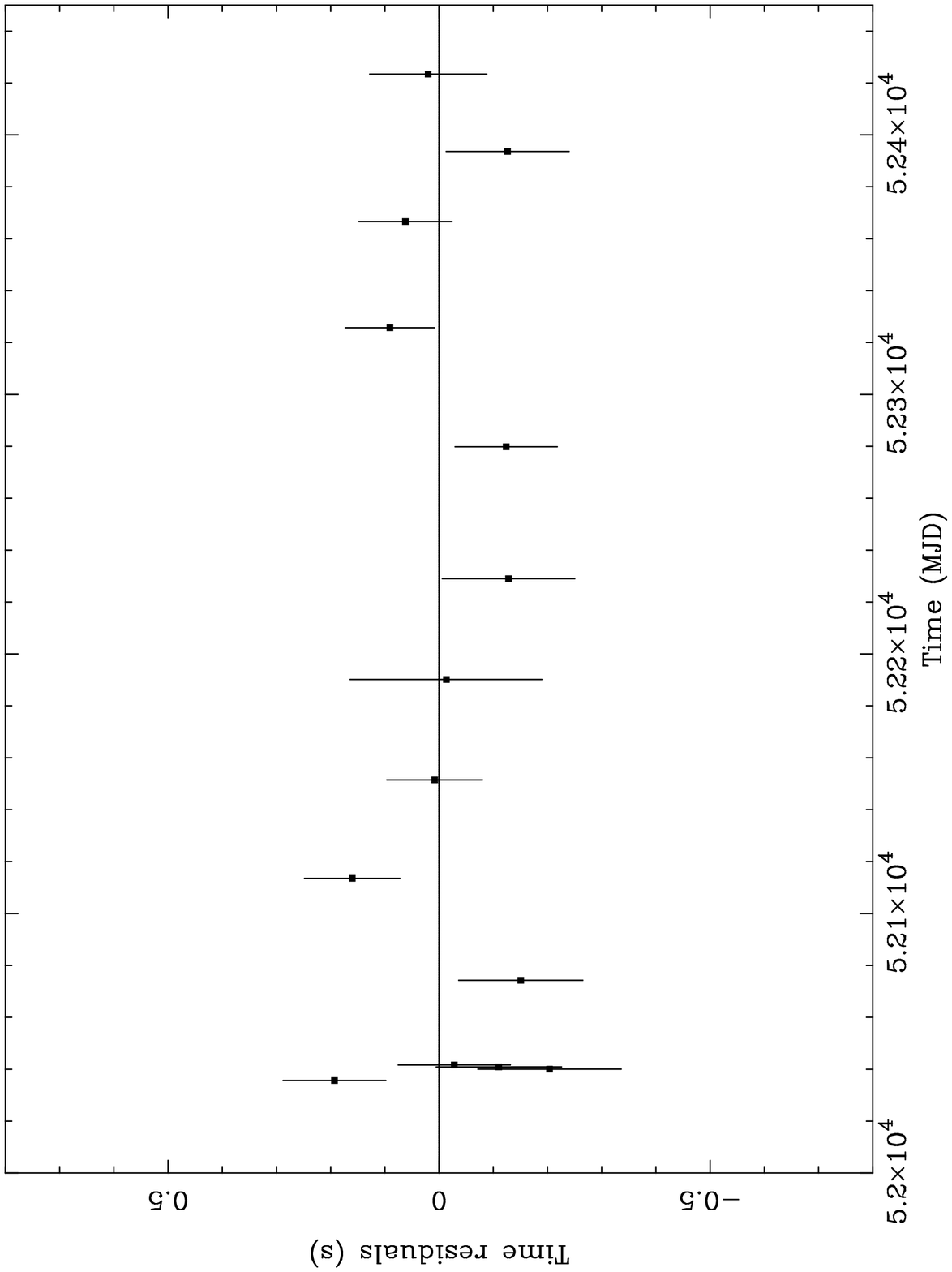}
\caption{}
\label{expres}
\end{figure*}

The plots of the best--fit solution of Eq. (\ref{esp}) to the phase 
residuals after subtraction of the preglitch model and of time 
residuals with respect to this best--fit solution are shown in Fig. \ref{exp} 
and Fig. \ref{expres} respectively.

In conclusion, the recovery from glitch N.2 is consistent with an exponential
decay with a timescale $\tau \sim 50$ d. There is evidence for a steepening 
($\simeq$ 0.9 \%) of the long--term spin--down rate as well, which is an order
of magnitude greater than observed permanent changes in Crab--like glitches
 while comparable to changes recovered linearly over a few years in Vela--like 
glitches.
Future observations will help revealing the possible long--term recovery of 
$\Delta \dot{\nu}_l$ (a significant $\ddot{\nu}$) or a possible 
$\Delta \nu_l$, which here are found to be both non significant.
                              
An exponential fit to the residuals of glitch N.1 after subtraction of the 
pre--glitch model required a further component superposed to the exponentially
recovered spin--up ($\tau \sim$ 350 d), either a negative step in the frequency
 ($\Delta \nu_l /\nu =(-3.9 \pm 0.7)\times 10^{-7}$) or an increase (in 
absolute value) of the first derivative ($\Delta \dot{\nu}_l/ \dot{\nu}=
(3.1\pm 0.6)\times 10^{-3}$).
In either case one obtains similarly good fits (respectively $\chi^2$=13.7 and 
13.47 with 15 d.o.f.), while including both components provides no significant 
improvement of the fit. Stated differently, at glitch N.1 data do not allow an 
unambiguous determination of the exponential recovery solution (Eq. \ref{esp}).
Thus the polynomial solution for this glitch is still preferred.

\section{Discussion}

\subsection{RXS J170849.0--400910 compared to radio pulsars}
\label{comparison}

As a first step in the investigation of our results we compare them with some
qualitative correlations between glitch variables studied by Lyne et al. 
(2000) in a sample of 48 radio pulsar glitches.

It is usually assumed that glitches are a manifestation of the angular momentum
 exchange between the bulk of a neutron star's moment of inertia and some 
superfluid component, loosely coupled to it, that slows down at discrete steps.
Let the spin--up rate associated with glitches be 
$\dot{\nu}_{glitch}=(\Sigma_{i} \Delta \nu_i)/ \Delta t$, where the summation 
is over all glitches observed during the time $\Delta t$. Lyne et al.(2000) 
found a good correlation between the long--term spin--down rate of radio 
pulsars and $\dot{\nu}_{glitch}$. Their best--fit linear relation is:

\begin{equation}
\label{Lyne}
\dot{\nu}_{glitch}= 1.7 \times 10^{-2} \dot{\nu}
\end{equation}

In the framework discussed above, conservation of total angular momentum and 
eq. (\ref{Lyne}) imply that a fraction $\sim 1.7 \times 10^{-2}$ of the moment 
of inertia of the star is on average contributed by the superfluid that is 
slowed down at glitches (see Sec. \ref{vortexcreep}).

In the case of the two glitches observed in RXS J170849.0--400910
$\dot{\nu}_{glitch} \sim 2.6 \times 10^{-15}$ s$^{\mbox{{\rc -2}}}$, which is 
strikingly close to the value (2.7 $\times 10^{-15} \mbox{s}^{\mbox{{\rc -2}}}$) predicted by formula (\ref{Lyne}). 
We stress however that this result still needs a confirmation, being based on 
only two events.

Further, the polynomial fit to the recovery from glitch N.1 is consistent with 
the trend observed in radio pulsars for rotators with decreasing $\dot{\nu}$ 
to have smaller exponential recovery fractions (Q) and increasingly important 
long--term linear recoveries. Based on such observed trend a recovery fraction 
Q $\sim 10^{-3}$ would be expected for RXS J170849.0--400910 
(se Fig.6 of Lyne et al. 2000).
A short--time transient of this magnitude cannot be excluded given
the sparse temporal monitoring of this source.

On the other hand, the trend is contradicted by the exponential fit to the 
recovery from glitch N.2. This is dominated by an exponential component
(Q $\sim$ 1). Recovery fractions greater than 10\% have been found only in the 
fastest and youngest (the Crab pulsar) and in the slowest and oldest 
(PSR B0525+21) radio pulsars in the sample of Lyne et al. (2000). AXPs do 
likely spin--down under different conditions with respect to common radio 
pulsars and may have higher temperatures than radio pulsars of the same 
spin--down age. Both effects may be  relevant in determining the properties of 
their recovery from glitches.

We notice that two other suggestions made by Lyne et al.(2000) do not fit with 
what is observed in RXS J170849.0--400910; the tendency for stars with a lower 
rotational frequency to experience, if any, glitches of smaller amplitude and, 
second, the tendency for stars with a higher dipole field to show more 
frequent but smaller glitches, at a given characteristic age. Indeed, the
average $\Delta{\nu}/{\nu}$ of the two observed glitches amounts to $\sim$
2 $\times 10^{-6}$ , very close to the value of the Vela pulsar 
($\sim 1.7 \times 10^{-6}$ as derived from Tab. 3 of Lyne et al. 2000). 

Thus, though the average amplitude of the glitches in 1RXS J170849.0--400910 
and the interglitch time are consistent with those found in radio pulsars of 
comparable spin--down age (particularly with those of the Vela pulsar), 
looking more closely at the results one finds peculiarities difficult to 
account for.

First, the average glitch ampitude and the short interglitch time are 
unexpected, since slow rotators do on average experience glitches of smaller 
amplitude and less frequently. This result may rule out a clear dependence on 
the rotation period and suggest that the main role could be played by the age 
of the neutron star. Indeed, contrary to radio pulsars with comparable 
rotational frequency, AXPs are slow but relatively young rotators. 

Second, a fractional increase of $\dot{\nu}$ as high as $\sim$ 30\%, and with 
a relatively long recovery timescale (tens to hundreds days) has never been 
observed in glitching radio pulsars. The only other source with such an 
anomalous behaviour is 1E 2259+586, in which the glitch recently detected 
produced a change of $\sim$ 100\% in the spin first derivative, recovered over
$\sim $60 d (Kaspi et al. 2003).
The fact that both glitching AXPs, and only them, have displayed such 
changes of $\dot{\nu}$ after a glitch hints to an AXP mechanism triggering
more extreme events than in radio pulsars, but a larger sample of 
glitches and of sources is needed to check this possibility.

Finally, one may explain the high fractional change of $\dot{\nu}$ 
if the recovery in glitch N.2 is modeled as an exponential. Even in this case, 
however, a problem is presented by the very smooth and linear recovery seen 
after glitch N.1; the star appears to have undergone an important change in its
 dynamical response over the short time interval elapsed between the two 
glitches. A way out may be assuming that the two events have been triggered by 
two different mechanisms, a possibility that will be discussed along with 
others in the following sections.  

\subsection{Vortex creep model applied to RXS J170849.0--400910}
\label{vortexcreep}

It is generally assumed that neutron--rich nuclei in the inner crust of a 
neutron star act as pinning centers for the vortices created by the superfluid
in response to the rotation of the star (Andersson \& Itoh 1975).

The superfluid can follow the crust's spin--down by outward migration of
vortices, which is impeded by the pinning energy barrier. 
In vortex creep model (Alpar et al. 1984, 1989, 1993) the rotational lag 
($\omega = \Omega_{sf} - \Omega_c$) that develops between the crust ({\rc 
spinning} at $\Omega_c$) and the superfluid (spinning at $\Omega_{sf}$) causes 
a Magnus force to arise, so that vortices overcoming the pinning energy 
barrier through thermal activation are driven outwards by this force, 
establishing a continuous vortex current through which the superfluid spins 
down. In doing so, it exerts an internal spin--up torque on the crust opposed 
to the magnetic braking.  
Eventually a steady--state is reached in the relaxation time $\tau$, when 
both the crust and the superfluid spin--down at the same rate 
($\dot{\Omega}_{\infty}$) with a steady--state lag 
$\omega_{\infty}$, subject to the external torque N$_{ext}$.

In steady--state, the equation of motion is simply:

\begin{equation}
\label{slowdown}
(I_c+I_{sf})\dot{\Omega}_{\infty} =
N_{ext}
\end{equation}
A glitch results from an instability of the steady--state. Inhomogeneities of 
the pinning force throughout the crust determine a non uniform distribution 
of vortices: local values of $\Omega_{sf}$ will be enhanced by accumulation 
of vortices until a critical lag $\omega_{cr}$ is reached, beyond which an 
avalanche unpinning of vortices occurs and angular momentum is transfered to 
the crust. Relaxation 
to the new steady--state will determine the post--glitch recovery.
The value of $\omega_{cr}$ determines whether the pinning is strong or weak. 
At the high density of the deep crust a change in the nature of pinning is 
thought to occurr, bringing it to a superweak regime, even though this 
conclusion is not testable in terrestrial materials. 

The conservation of total angular momentum during the glitch gives
(Alpar et. 1993, Alpar 2000):

\begin{equation}
\label{deltaomega}
(\alpha I_A + I_B) \delta \overline{\Omega} = I_c \Delta \Omega_c
\end{equation}
where $I_A+I_B=I_{sf}$ is the moment of inertia of the pinned superfluid, 
$\delta \overline{\Omega}$ is its change in rotational frequency
and the right hand side refers to the crust plus all matter 
components coupled to it on very short timescales: these include the 
bulk of the core material (Alpar, Langer and Sauls 1984) which makes up most of the 
moment of inertia of the star.
Two different situations may occur in the crustal superfluid:
in type--A (accumulation) regions unpinning takes place, vortices move outwards
 and repin in another type--A region. For a uniform density of unpinned 
vortices $\alpha=\frac{1}{2}$ ($0< \alpha <1$ in general).

Type--B regions exist between accumulation zones and are vortex--depleted 
regions. Thus they are decoupled from the crust and spin--down only at 
glitches, when a net outward flux of vortices crosses them. Thus they 
contribute to $\Delta \Omega_c$ at the glitch but not to changes in 
$\dot{\Omega}_c$.

The relaxation time $\tau$ depends on whether the response of the superfluid
to the initial perturbation of the lag, $\delta \omega(t=0)$, is linear or
non linear. The corresponding expressions are (Alpar et al.1989):

\begin{equation}
\label{taulin}
\tau_l= \frac{kT}{E_p} \frac{\omega_{cr} r}{4 \Omega v_0} \mbox{exp}(\frac{E_p}{kT})
\end{equation}
\begin{equation}
\label{taunonlin}
\tau_{nl}= \frac{kT}{E_p} \frac{\omega_{cr}}{|\dot{\Omega}|_{\infty}}
\end{equation}
in the linear and non linear regime, respectively.
The former situation will occur when $\omega_{\infty}\ll \omega_{cr}$,
the latter when the steady--state is close to the condition for unpinning, 
$\omega_{\infty}\leq \omega_{cr}$.
The system will always be in the regime with the shortest time, so a 
transition value for $(E_p/kT)$ can be found by equating the two 
(Alpar et al. 1993). For RXS J170849.0--400910 it obtains 
$(E_p/kT)_{cr}\simeq 30.8$.
If the ratio is below this limit the superfluid responds linearly, while 
non--linearity corresponds to a value higher than critical. 

The non--steady--state equivalent of eq.(\ref{slowdown}) in the linear and non 
linear regime are, respectively (Alpar et al. 1989):

\begin{equation}
\label{linear}
\dot{\Omega}_c = \dot{\Omega}_{\infty} - \frac{I_{sf}}{I_c} \frac{\delta \omega(0)}{\tau_l} \mbox{exp} (- \frac{t}{\tau_l})
\end{equation}

and 

\begin{equation}
\label{nonlin}
\dot{\Omega}_c = \frac{I}{I_c}\dot{\Omega}_{\infty} - \frac {I_{sf}}{I_c} 
\dot{\Omega}_{\infty} \frac{1}{1+[\mbox{exp} (t_0 / \tau_{nl}) - 1]\mbox{exp} (-t / \tau_{nl})}
\end{equation}
where $t_0 =\frac{\delta \omega(0)}{|\dot{\Omega}|_{\infty}}$ is an offset 
time indicating how long, after the decoupling, the region of interest will 
start recoupling through creep. 

In the linear regime an exponential recovery is expected while nonlinear
regions have a more complex response. With $\alpha$=(1/2) and $\tau_{nl} < 
(t_0)_{max} =\frac{\delta \overline{\Omega}}{|\dot{\Omega}|_{\infty}}$, 
$\dot{\Omega}_c$ will recover linearly in time due to the progressive 
recoupling of type--A regions. 

Glitch N.1 has been previously interpreted in terms of a non--linear response. 
If most of the superfluid had been involved in vortex motion 
(then $I_B \ll I_A$), a fraction 1.72$\times 10^{-2}$ of
the whole moment of inertia of the star needs to be contributed by the pinned 
superfluid.
The expected time to the next glitch is then $t_g\sim (t_0)_{max} \simeq 500$ 
days (a smaller value ensues if allowance for a significant $I_B$ is made), 
remarkably close to the measured interval of 550 days. For the recovery to be 
linear in time it is required that $\tau_{nl} < t_g$, implying 
$\omega_{cr} < 4.3\times 10^{-5} \left(E_p/kT\right) < 10^{-2} 
\mbox{rad s}^{-1}$ even for the largest values of the ratio ($E_p/kT$).

From Eq.(\ref{deltaomega}) and the glitch amplitude in Tab.\ref{glitch1} one 
gets $\delta \overline{\Omega} \leq 4.2\times 10^{-5} \mbox{rad s}^{-1}$. 
The condition (Alpar et al. 1989) $\omega_{cr} > \delta \omega \simeq \delta 
\overline \Omega$ for the non--linear regime thus gives the lower bound 
$\omega_{cr} > 4\times 10^{-5} \mbox{rad s}^{-1}$. A non linear regime obtains 
only for a very low value of $\omega_{cr}$, requiring superweak pinning 
throughout most of the crust. This is surprising since superweak pinning 
should characterize the innermost and densest layers of the crust.

The change in $\dot{\nu}$ measured in glitch N.2 cannot arise from a 
non--linear response. The fractional change in $\dot{\nu}$ in this regime is 
always less than the ratio $(I_{sf}/I_c)$ (Eq.\ref{nonlin}), thus an 
unplausibly high moment of inertia of the superfluid would be implied.
Only the linear response (and consequently an exponential recovery) of a 
region where unpinning did take place ($\delta \omega \simeq \delta 
\overline {\Omega}$) can account for such a high fractional change. If the two 
glitches were due to the same mechanism, the recovery from glitch N.1 should 
be modeled as an exponential as well or one should invoke a change in the 
response regime of the superfluid.

In the case of two exponential recoveries, the different timescales hint to 
the association of the two glitches with different regions of the superfluid. 
In particular, they must reflect the different values of $\omega_{cr}$ or 
$E_p$. The value of $\tau_l$ is very sensitive to the ratio ($E_p/kT$) 
(Eq.\ref{taulin}), so that a difference of a factor $\sim6$ can arise 
from a small difference in that ratio. Since both recoveries are fit
with only one of the two exponential components, these must have unpinned
alternately. Stated differently, the possibility of local unpinning should be 
introduced in the model. 

The shorter recovery timescale of glitch N.2 would require a smaller 
$\omega_{cr}$ or a lower $(E_p/kT)$, corresponding to a region of weaker 
pinning, given the same response regime. In particular, the smaller value of 
$\omega_{cr}$ would imply a correspondingly larger fraction of superfluid to 
have unpinned, in order to produce the larger glitch N.2.

Glitch N.1 may have originated in the outer layers of the crust, where
$\omega_{cr}$ is likely higher, leaving the more weakly pinned, inner 
superfluid unperturbed (apart from the very small $\Delta \Omega_c \sim 4
\times 10^{-7} \mbox{ rad s}^{-1}$). The roles of the two components should be 
interchanged to explain glitch N.2. Anyway, in vortex creep model unpinning is 
a widespread rather than local mechanism (Pines 1999, Jones 2002), so this 
would be an ad hoc assumption.
 
If one modeled the recovery from glitch N.1 as a linear decay, a change 
in the response regime of at least some parts of the superfluid must have
 occurred over the $\sim$ 500 days of interglitch time. This may happen if
after glitch N.1, where the response was non linear, a significant release of 
heat acted to decrease the ratio $E_p/kT$, bringing the latter below the 
transition value. However, such a heat release would require some new 
mechanism not included in vortex creep model (\textit{e.g.} a starquake, 
see Sec. \ref{starquake}).

Furthermore, as already discussed in Sec. \ref{nuovidati}, the period evolution
after glitch N.2 is unusual. After the initial recovery of the spin--up, the 
spin frequency becomes lower than expected from the extrapolation of the 
pre--glitch solution. In the exponential fit this effect is due to the 
unrecovered increase in the long--term spin--down rate.
Link, Epstein and Baym (1992) have shown that in vortex creep model, where no 
explicit temporal dependence of the net (external+internal) torque acting on 
the crust is included, the post--glitch spin frequency must always remain 
higher than the extrapolation of the pre--glitch value, approaching it in the 
long run. Thus Fig. \ref{aprilglitch} argues against an interpretation of 
glitch N.2 in terms of vortex creep model, unless allowing for a change in 
the torque, either external or internal, induced by the glitch.  

In summary, one needs a contrived interpretation in order to account for the 
different amplitudes and recoveries of the two observed glitches within the
framework of vortex creep model.
In fact one should assume the two glitches to be produced by different 
mechanisms, thus attributing the different recoveries to the response of the 
star to different kinds of perturbation.

\subsection{Core superfluid models}
\label{core}

One of the basic assumptions of crustal superfluid models is that the core 
neutron superfluid be strictly coupled to 'the protons' (the star's crust, 
core superconducting protons and degenerate electrons) on the 
timescales of post--glitch recoveries. The coupling is provided 
by scattering of the electrons -- following the crust spin -- off 
magnetic flux tubes that thread neutron vortex cores as a result of the 
entrainment--induced circulation of superconducting protons around them 
(Alpar, Langer and Sauls 1984, Prix et al. 2002). 
It has been suggested (Andersson and Comer 2001, Andersson et al. 2002, 
Sedrakian \& Sedrakian 1995) that the above assumption may be questionable, as
the coupling time $\tau_c \propto(c_r+c^{-1}_r)$ between neutron vortices and 
'protons' depends on the uncertain parameter $c_r$, the ('drag--to--lift') 
ratio between the drag force due to entrainment and the Magnus force, both 
acting on neutron vortex lines in the decelerating core. 
From the above expression, $\tau_c$ is found to be smallest for $c_r\sim 1$,
 while in both cases of weak ($c_r\ll1$) and strong ($c_r\gg1$) coupling, 
$\tau_c$ can increase significantly.

Andersson et al. (2002) studied a two--superfluid instability that may take 
place in the core of neutron stars if a rotational lag between the superfluid 
neutrons and 'the protons' can develop, due to a long coupling time. The 
instability triggers a sudden rearrangement of the neutron vortex distribution 
in an outer core shell, leading to a glitch.
In this model a simple formula is given to estimate the interglitch time
$t_g$ in RXS J170849.0--400910. The instability takes place when the 
rotational lag ($\omega$) between the two components is large enough 
to overcome the coupling due to the drag force ($\omega > \omega_{cr}$). Thus:

\begin{equation}
\label{interglitch}
t_g= 2 t_s \frac{\omega_{cr}}{\Omega_c}
\end{equation}
where $\omega_{cr}$ has an analogous meaning as it has in vortex creep model
while $t_s$ is the spin--down age of the star.

Assuming $\omega_{cr}\simeq 5\times 10^{-4}\mbox{rad s}^{-1}$ as a
typical value of the lag for unpinning (Andersson et al. 2002 and references
therein), no glitch would be expected within $\sim$ 30 years after glitch N.1. 
Thus, for this model to apply to RXS J170849.0--400910, a very low value of 
$\omega_{cr}\leq 3\times 10^{-5} \mbox{rad s}^{-1}$ would be required.  
Whether such a low value still allows the lag to effectively develop is an 
open question.

An attractive feature is that, if 30 yrs did represent the characteristic 
recurrence time for the core instability in 1RXS J170849.0--400910, this model
would allow a natural interpretation of the different properties of the two 
observed glitches, since these could not be due to the same mechanism.
In particular, if glitch N.2 was due to this instability, a substantial 
fraction of the core superfluid could have decoupled, thus explaining the 
large increase of $\dot{\nu}$ as a significant reduction in the moment of 
inertia acted on by the braking torque. 
The different responses and recovery timescales of the two glitches would be 
-- at least qualitatively -- accounted for, since in glitch N.2 they would 
reflect mainly the dynamical relaxation of the core rather than crustal 
superfluid.

Thus two different glitch mechanisms could be at work in the same source. 
Future timing observations will help testing this interesting possibility, as 
more glitches are expected. Since it may take more than a decade for a 
new event similar to glitch N.2 to occur, the lack of such events in the next 
years may indeed provide an indirect hint to the viability of this picture.

\subsection{The starquake scenario}
\label{starquake}

It has been recently shown (Wang et al. 2000) that most glitch models are 
difficult to reconcile with observations of a growing number of glitches. 
All models requiring a catastrophic (\textit {i.e.} widespread) unpinning of 
crustal superfluid require simple correlations between the amplitudes and 
recovery timescales of glitches in a single source as well as an approximately 
constant Q--value, which is found not to be the case, in general.
This is the same difficulty met in Sec. \ref{vortexcreep} for the two 
glitches of 1RXS J170849.0--400910. We suggested the possibility that 
these events were due to a local, rather than global, release of angular 
momentum. This is the same conclusion of Wang et al.(2000) based on a sample 
of 30 glitches.

An entirely different explanation with respect to those discussed previously 
has been proposed for at least the large glitches of radio pulsars: these may 
be associated to starquakes and subsequent movements of cracked platelets 
(Ruderman 1991, Ruderman, Zhu and Chen 1998, Epstein and Link 2000). 
Several cracking mechanisms have been proposed for radio pulsars; quakes have 
been ascribed to the stress acted on the crust by the interaction in the core 
between superfluid neutrons and magnetic flux tubes threading the crust 
(Ruderman, Zhu and Chen 1998). Alternatively, the growing strain in the crust 
of a spinning--down neutron star may fracture it due to the resulting shear 
stress (Franco, Link and Epstein 2000). 
Whatever the trigger, the movement of a sector of the crust with its 
frozen--in magnetic flux can cause an increase of the angle $\alpha$ between 
the rotation and magnetic axis and a subsequent increase of the braking torque 
exerted on the star.

The most important feature of this scenario is its local nature; only the
moving sector and its surroundings are affected leaving the rest of the star 
mainly unperturbed (Jones 2002).
Further, following a starquake, a permanent increase of $\dot{\nu}$ could 
result as a consequence of the net increase in the braking torque. A residual 
torque increase has indeed been observed after several glitches in the Crab 
pulsar which may result from permanent changes in the magnetic torque 
(Link, Epstein and Baym 1992, Link \& Epstein 1997).
Our exponential fit to the recovery from glitch N.2 provides evidence for an 
increase in the spin--down rate, though greater than those observed in the 
Crab pulsar. Whether this increase is recovered over a timescale longer
than 1 yr or not is a crucial question to be answered only with future 
observations. In vortex creep model this change would be attributed to some of 
the pinning layers responding non linearly to the glitch and recoverying over 
a time $t_g=\delta \omega / |\dot{\Omega}|_{\infty}$. If vortices did unpin in 
these layers a linear recovery of $\dot{\nu}$ is expected, while if they did 
not unpin $\delta \omega=\Delta \Omega_c$ and a step--like (Alpar et al. 1984) 
recovery of $\dot{\nu}$ would occur, after a waiting time 
$t_0=\Delta \Omega_c / |\dot{\Omega}|_{\infty}$.
On the other hand, finding no significant recovery of $\Delta \dot{\nu}_l$
in future observations would be in agreement with the expectations from 
plate tectonics.

In the model by Ruderman, Zhu and Chen (1998) glitch activity is expected to 
cease at rotational periods P$>$ 0.7 s because of the decreasing pull of core 
neutron vortices on flux tubes. Thus an intrinsically different mechanism 
would be required to cause starquakes in slow rotators such as AXPs.
In the case of shear stress--induced fractures, the high spin--down of AXPs
may well produce starquakes. Alhough it has been proved to be impossible on 
energetic grounds to account for the frequent glitches (every $\sim$ 2 yrs) 
of the Vela pulsar through this mechanism (Sauls 1988 and references therein), 
we stress that a Vela--like glitch in an AXP ($\Delta \nu/\nu \sim 10^{-6}$) 
will involve some 4 orders of magnitude less energy ($\Delta E \propto \nu^2$).
 Thus the energetic argument may still leave room for spin--down--induced
glitches in AXPs. 
However AXPs in the magnetar model are candidates for starquakes triggered by 
a peculiar mechanism (Thompson and Duncan 1996). The super strong magnetic 
field diffusing through the core can induce stresses in the crustal lattice 
strong enough to crack it. A range of length scales for magnetically--driven 
fractures is expected; continuous small scale fracturing is produced by the 
diffusion of the field through the crust while larger scales, comparable to 
the thickness of the crust, are likely involved by sudden rearrangements of 
the field distribution.
When the crust yields, the sudden shaking of magnetic footpoints excites 
magnetospheric Alfv\`en waves; in particular, burst--like enhancements of the 
X--ray emission are expected in association with larger scale fractures. This 
mechanism has been proposed to explain the repeat bursts of Soft Gamma 
Repeaters (Thompson $\&$ Duncan 1993) and may be relevant also to AXPs. 
Indeed, the simultaneous occurrence of a major X--ray outburst and a large 
glitch in the AXP 1E2259+86 (Kaspi et al. 2003) has provided a strong hint in 
favour of this interpretation. On the other hand, observations of burst 
activity in SGRs have revealed no clear correlation with glitch--like changes 
in the rotational parameters of the sources (Woods et al. 2002, 2003), making 
the interpretation of such effects still controversial. 

Though the mechanism for cracking the crust may be different in AXPs and radio 
pulsars, the post--starquake evolution is likely to be similar, if determined 
mainly by the dynamical coupling of the superfluid components to crustal 
matter. Jones (2002) has investigated the post--glitch evolution of a neutron
star following a starquake, taking into account the responses of both 
superfluids in the crust and the core. In that picture several physical 
parameters determine the mutual interaction between the superfluid components 
and the magnetic field and the interaction of crustal vortex lines with 
crustal nuclei. The location at which the fracture occurs is a key element, 
determining the relative importance of such parameters.
A variety of glitch amplitudes, recovery fractions and timescales is expected 
not only in different sources but also in a single one, as fractures with 
different dimensions and at different depths would occurr.

As a general remark, the complexity of the model and its dependence on poorly 
known parameters make predictions, if any, and quantitative observational tests
very difficult. On the other hand, its complex nature makes it more suited 
than other models to the description of a variety of different glitches. 
In particular, the two glitches of RXS J170849.0--400910 do require a complex 
picture -- if not ad hoc assumptions -- even within other models. 
In the starquake scenario the two glitches would be explained assuming that 
the cracking and platelet movement did happen at two different locations 
within the crust. The unrecovered increase in the spin--down rate ($\leq$ 1\%) 
of glitch N.2 could then be attributed to a permanent change in the braking 
torque. This would require a change in the inclination angle between the 
magnetic and rotation axes $\Delta \alpha \leq 10^{-2} \mbox{tan} \alpha$, 
actually a large value compared to expected ones ($\leq 10^{-3}$, Ruderman, 
Zhu and Chen 1998).
Finally, a change in the magnetic field orientation would be expected to 
produce some changes in the shape of the pulse profile (Franco, Link and 
Epstein 2000), such as those we may have found here, particularly after 
glitch N.2.

\section{Conclusions}

We have obtained a timing solution for 1RXS J70849.0--400910 spanning more 
than 4 years. We detected a new large glitch, whose characteristcs place this 
source among the most frequent glitchers, undergoing large amplitude 
spin--ups.

The two glitches are perhaps an indication of yet another peculiarity of AXPs.
Indeed, the source has experienced two large glitches followed by very 
different recoveries, the larger being recovered more quickly.

Our results allow for the first time a comparison between the glitch 
activity of an AXP and that of radio pulsars, an important step in the 
assessment of the nature of AXPs as extremely magnetised, isolated neutron 
stars. Our observations show that:

\begin{enumerate}

\item  The short interval ($t_g<$ 2 years) between the two glitches of 
1RXS J170849.0--400910 is comparable to the intervals found in the Vela pulsar,
 one of the most frequent glitchers. The two observed spin--ups are large, 
with average amplitude $\Delta \nu / \nu \sim 2\times 10^{-6}$ comparable to 
those of Vela and other so--called 'adolescent' pulsars (Lyne et al. 2000).

\item  The average glitch--induced spin--up torque is comparable to that of 
the Vela pulsar and consistent with that found in radio pulsars of age 
$t_s>10^4$ yrs and with high enough $\dot{\Omega}$ (Lyne et al. 2000).

\item The large average glitch amplitude ($\sim 2\times 10^{-6}$) is at 
variance with what observed in slow rotators, that appear on average to 
suffer rarer and smaller glitches. 
If one assumes AXPs to be magnetars, the large amplitude of the glitches is 
also at variance with the tendency for higher magnetic field neutron stars to 
experience smaller (but more frequent) glitches, at a given age (Lyne et al.
2000).

\item  The two glitches are characterized by different types of recovery.
 The change in the rotational parameters after the second glitch is similar, 
though less extreme, than the recently detected burst--associated glitch in 
the AXP 1E 2259+586 (Kaspi et al. 2003).

\item  There is evidence that the second glitch consisted of a large
spin--up recovered exponentially over $\sim$ 50 d plus a steepening
of the spin--down rate of the order of $\sim$ 0.9\%. Further components, 
though non significant in the fits yet, cannot be ruled out.
A similar interpretation may hold for glitch 1 even though, in this
case, data are wholly consistent with a simple linear recovery. 

\end{enumerate}
The two different recoveries are puzzling. No simple model can account for this
finding and ad hoc assumptions would be required.

Observations do not seem to fit with models based only on instabilities of the
crustal superfluid, such as vortex creep model. Indeed the recoveries have 
variable parameters from one glitch to the other, which would either require 
unpinning to occur in separate localized regions at each glitch or changes in
the response regime of parts of the superfluid. Further, while an 
interpretation of glitch N.1 in terms of a non linear response within vortex 
creep model is perfectly consistent with observations, new hypotheses should 
be introduced in order to explain the decrease, observed after glitch N.2, of 
the rotational frequency below the extrapolation of the pre--glitch value. 

A likely picture is that the two glitches were triggered by two different 
mechanisms. Models invoking a perturbation in the core superfluid may fit the 
properties of glitch N.2, in particular the inferred large change of 
$\dot{\nu}$. Further, in this case the long expected interglitch time would 
naturally lead to the conclusion that glitch N.1 could not be due to the same 
mechanism and provide a key to understanding the two different recoveries.

However, starquake models seem more promising. In general, their local nature
and the complex picture of the glitch phenomenon that they provide seem to fit 
well with observations of a great variety of glitches in radio pulsars.
Concerning AXPs, the glitch recently detected in 1E2259+586 is similar to 
glitch N.2 of 1RXS J170849.0--400910, having a comparable magnitude in 
$\Delta \nu$ and an even greater change in $\dot{\nu}$. Put together, these 
two events may suggest that a peculiar glitch mechanism be at work in AXPs. 
In the magnetar model this might be associated to crustal cracking and/or 
plastic flow induced by the diffusing magnetic field. Indeed, the simultaneous
outburst and glitch of 1E 2259+586 do provide a first strong hint that crust 
cracking--induced glitches can occur in AXPs, reinforcing a similar 
interpretation at least for glitch N.2 observed here, given their overall 
similarity. Though we found no evidence for bursts in 1RXS 
J170849.0--400910, our sensitivity to very short and hard events was limited. 
Moreover, both glitches did take place during $\sim$ 40d time intervals 
between subsequent observations, so no firm conclusions about this point can 
be drawn.
Finally, since no clear correlation between glitch and burst activity has been 
found in SGRs, it seems that observing them together may be the exception 
rather than the rule.   

In the case of 1RXS J170849.0--400910, a starquake interpretation can account 
for the observed decrease, at glitch N.2, of the rotational frequency of the 
source below the extrapolation of the pre--glitch solution and also for the 
small, though significant, changes in the average shape of the pulse profile, 
particularly after glitch N.2.
On the other hand, starquake models involve many degrees of freedom and have, 
up to now, poor predictive power. More developments in modeling these events 
are needed in order to closely match them with observations. Further, future
observations are needed in order to get a clearer picture of the glitch 
phenomenon in AXPs.

\acknowledgements
~
\newline
This work was partially supported by University of Rome ``La Sapienza''. 
~
\newline
S.D. would like to thank the anonymous referee for helpful comments to the 
paper.

\vfill \eject

\vfill\eject
\newpage

\section*{Figure Captions} 

\noindent {\large \bf Figure 1 --- } Time residuals of the pre--glitch model 
for observations between 1998 January 12 (MJD 50825) and 1999 September 25 
(MJD 51446)
 
\noindent {\large \bf Figure 2 --- } Time residuals of observations from 1999 
October 21 (MJD 51472) to 2001 May 29 (MJD 52423) after subtraction of the 
post--glitch model N.1. The sharp increase (in absolute value) of the 
residuals between 2001 March 27(MJD 51995) and 2001 May 6 (MJD 52035) 
indicates the occurrence of glitch N.2. The vertical line corresponds to the 
assumed epoch of the glitch.

\noindent {\large \bf Figure 3 ---} Time residuals of observations from 1999 
October 21 (MJD 51472) to 2002 May 29 (MJD 52423) after subtraction of 
post--glitch model N.1 previous to May 2001 and the polynomial post--glitch 
model N.2 after (see Tab.\ref{glitch2} for parameters). The vertical line 
indicates the assumed epoch of the glitch.

\noindent {\large \bf Figure 4 ---} Average pulse--profiles for the three intervals of Tab. \ref{pulsemedio}, from top to bottom Interval 1, 2 and 3.

\noindent {\large \bf Figure 5 ---} Phase--residuals of the observations from 
2001 May 6 to 2002 May 29 relative to the preglitch model (left column in 
Tab.\ref{glitch2}). The 0 in the abscissa is the assumed epoch of the glitch
(MJD 52015.65).

\noindent {\large \bf Figure 6 ---} Time residuals of the recovery from glitch 
N.2 (2001 May 6 -- 2002 May 29) relative to the best--fit exponential+linear 
term of Eq. \ref{esp}.

\vfill\eject
\newpage

\newpage

\vfill \eject
\newpage

\begin{table*}
\begin{center}
\caption{Measured rotational parameters of 1RXS J170849.0--400910 previous and 
after glitch N.1 and inferred glitch parameters. 1$\sigma$ errors in the last 
digit are quoted in parenthesis.}
\label{glitch1}
\begin{tabular}{lcc}
\tableline
\tableline
Spin Parameter & Pre--glitch N.1 &  Post--glitch N.1\\
\tableline
~&~&~\\
Assumed     &         &         \\
glitch epoch (MJD) & 51459.0 & 51459.0 \\
$\nu$ (Hz) & 0.0909136405(1) & .090913699(2) \\
$\dot{\nu}$ ($\times 10^{-13} s^{-1}$) & -1.56809(12)   & -1.595(2) \\
$\ddot{\nu}\tablenotemark{a}\times 10^{-23} s^{-3}$ & - & 4.9(6) \\
MJD range & 50825--51446 & 51472--51995 \\
N. of fitted points & 39 & 19 \\
r.m.s. (s) & 0.16 & 0.09  \\
~ & & \\
$\frac{\Delta \nu}{\nu} \times 10^{-7}$ & \multicolumn{2}{c}{6.43(5)} \\
~ & & \\
$\frac{\Delta \dot{\nu}}{\dot{\nu}} \times 10^{-2}$ & \multicolumn{2}{c}{1.72(5)} \\
\tableline
\end{tabular}
\tablenotetext{a}{This parameter was found to be statistically non significant
at the 99 \% level in the fit to observed phases previous to glitch N.1. The 
corresponding values of $\nu$ and $\dot{\nu}$ are those derived by the 
best--fit model to the observed phases with $\ddot{\nu}$ fixed at zero.} 
\end{center}
\end{table*}

\begin{table*}
\begin{center}
\caption{Measured spin parameters and inferred glitch parameters for glitch 
N.2. 1$\sigma$ errors in the last digit are quoted in parenthesis}
\label{glitch2}
\begin{small}
\begin{tabular}{lcc}
\tableline
\tableline
Spin Parameter & Pre--glitch N.2 &  Post--glitch N.2\\
\tableline
~&~&~\\
Ref. Epoch(MJD) & 51555.733 & 52041.254 \\
$\nu$ (Hz) & 0.0909123674(6) & 0.090905939(5)\\
$\dot{\nu}$ ($\times 10^{-13} s^{-2}$) & -1.591(1) & -1.964(26) \\
$\ddot{\nu} ~(\times 10^{-23} s^{-3})$  & 4.9(6) & 527(65) \\
$\nu^{(3)}~(\times 10^{-28} s^{-4})$ & - & -4.8(9) \\
$\nu^{(4)}~\times 10^{-35} s^{-5} $ & - & 2.1(5) \\
MJD range & 51472--51995 & 52035--52423 \\
N. of fitted points & 19 & 14 \\
\hline
\hline
Polynomial Model\\
\hline
Assumed            &          &       \\
glitch epoch (MJD) & 52015.65 & 52015.65 \\        
$\nu$ (Hz) & 0.090906084(14) & 0.090906387(13)\\
$\dot{\nu}~(\times 10^{-13} s^{-2})$ & -1.571(6) & -2.09(4)\\
$\ddot{\nu}~(\times 10^{-23} s^{-3})$  & 4.9(6) & 638(85) \\
$\nu^{(3)}~(\times 10^{-28} s^{-4}) $ & - &  -4.8(10) \\
$\nu^{(4)}~\times (10^{-35}) s^{-5} $ & - & 2.1(5) \\ 
r.m.s. (s) & 0.09 & 0.09 \\
~ & & \\
$\Delta \nu / \nu~(\times 10^{-7})$ & \multicolumn{2}{c}{33(2)} \\
~ & & \\
$\Delta \dot{\nu} / \dot{\nu}~(\times 10^{-2})$ & \multicolumn{2}{c}{33(3) } \\
\tableline
\tableline
Exponential Model\\
\tableline
Assumed            &            &           \\     
glitch epoch (MJD) &  \multicolumn{2}{c}{52015.65} \\
$\delta \nu_{exp} (\times 10^{-9}$ Hz) & \multicolumn{2}{c}{357(13)}\\
$ \tau~(\mbox{days}) $  & \multicolumn{2}{c}{50(2)} \\
$ \Delta \nu_l\tablenotemark{a}~(\times 10^{-8}\mbox{s}^{-1}) $ & \multicolumn{2}{c}{(-4.1$\div$1.5)} \\
$ \Delta \dot{\nu}_l\tablenotemark{b}~(\times 10^{-15} \mbox{s}^{-1})$ & \multicolumn{2}{c}{-1.382(27)} \\
r.m.s. (s) & \multicolumn{2}{c}{ 0.12} \\
\tableline
\end{tabular}
\tablenotetext{a}{This parameter was statistically non significant at the 99\%
confidence level in the fits and was then set equal to zero. 
A 99.9\% confidence interval is reported in parenthesis}
\tablenotetext{b}{The value and uncertainty of this parameter are determined 
with the non significant term $\Delta \nu_l$ fixed at zero.}
\end{small}
\end{center}
\end{table*}

\begin{table*}
\begin{center}
\caption{Best--fit parameters of the average pulse profiles, obtained after 
folding the observations in the following three intervals: 
1) 12 January 1998 (MJD 50825) -- 25 September 1999 (MJD 51446); 
2) 21 October 1999 (MJD 51472) -- 2000 June 19 (MJD 51822); 
3) 6 May 2001 (MJD 52035) -- 2001 30 August (MJD 52151). 
1$\sigma$ errors in the last digit are quoted in parenthesis}
\label{pulsemedio}
\begin{small}
\begin{tabular}{cccc}
\tableline
\tableline
Parameter & Interval 1 & Interval 2 & Interval 3\\
~ & ~ & ~ & ~ \\
\tableline
N. of observations   &  39  & 11  &   7\\
$A_1 (\times 10^{-2})$ & 8.09(8) &  6.1(1)  & 5.1(2)  \\
$A_2/A_1$ & 0.28(1)   &  0.29(2)  & 0.32(3)  \\
$A_3/A_1$ & 0.15(1)   &  0.22(2)  & 0.26(3)  \\
$A_4/A_1$ &  --       &  --       & 0.15(3)  \\
$\psi_1$  &  0.229(2)  & 0.246(3) & 0.230(5)   \\
$(\psi_2-\psi_1) \times 10^{-2}$ & 7.5(3)    & 4.3(6)   & 4(1) \\
$\psi_3-\psi_1$                  & 0.250(4)  & 0.227(5) & 0.228(9)\\
$\psi_4-\psi_1$                  &  --  &  --  & 0.30(1)\\
\tableline
\end{tabular}
\end{small}
\end{center}
\end{table*}

\end{document}